\documentclass[seceq]{ptptex}

\usepackage{graphicx}
\newcommand{\be}{\begin{equation}}
\newcommand{\ee}{\end{equation}}
\newcommand{\bea}{\begin{eqnarray}}
\newcommand{\eea}{\end{eqnarray}}
\newcommand{\nn}{\nonumber}

\newcommand{\spur}[1]{\not\! #1 \,}

\title{
Puzzles in charm spectroscopy%
}

\author{
P. \textsc{Colangelo},\footnote{ e-mail address:
Pietro.Colangelo@ba.infn.it}  
F. \textsc{De Fazio},
R.  \textsc{Ferrandes}  and
S.  \textsc{Nicotri}
}

\inst{
INFN - Sezione di Bari, via Orabona 4, I-70126 Bari Italy
}

\abst{%
\noindent We briefly  analyze  aspects of  open and hidden charm  resonances, discussing in particular the mesons  $D_{sJ}(2860)$ and $X(3872)$.}

\begin{document}

\maketitle

\section{Prologue}
The word {\it puzzle}  means   a {\it problem}, a mystery deserving explanation.  It also indicates a {\it game designed for testing ingenuity}, where  pieces of information have to be put together to reassemble  a known picture. 
 It is worth asking if  recent   results  in  charm spectroscopy~\cite{reviews}  
represent problems  or  information   fitting  into a known theoretical scheme.  The answer is different in case of open and hidden charm mesons.
\section{$c \bar s$ system and   $D_{sJ}(2860)$ }
An example of new experimental information fitting into an established theoretical scheme is the
 meson  $D_{sJ}(2860)$  recently observed by BaBar Collaboration~\cite{palano06} in the $DK$ system inclusively produced in  $e^+ e^- \to DKX$,   with
$M(D_{sJ}(2860))=2856.6 \pm 1.5 \pm 5.0$ MeV  and
$\Gamma(D_{sJ}(2860) \to DK)= 47\pm 7 \pm10$ MeV 
($DK=D^0 K^+ + D^+ K_S^0$).
Together with this state,
a broad structure was  noticed  with  $M=2688\pm4\pm3$ MeV and $\Gamma=112\pm7\pm36$ MeV;  indeed,  Belle Collaboration~\cite{belle2715}  reported the evidence of $D_{sJ}(2715)$   in $B^+\to \bar D^0 D^0 K^+$ decays,    with 
$M(D_{sJ}(2715))=2715\pm11^{+11}_{-14}$ MeV,
$\Gamma(D_{sJ}(2715)=115\pm 20^{+36}_{-32}$  MeV  and $J^P=1^-$.

The  interpretation of  these charmed resonances  is  easier in  the heavy quark limit $m_Q \to \infty$.
 In such a
limit  the spin $s_Q$ of the heavy quark  and  the angular momentum
 $s_\ell$ of the meson light degrees of freedom: $s_\ell=s_{\bar q}+ \ell$ ($s_{\bar q}$ 
 light antiquark spin,  $\ell$  orbital angular
momentum of the light degrees of freedom relative to the heavy quark) are decoupled, and the spin-parity  $s_\ell^P$  is conserved in strong interaction processes
\cite{HQET}.  Mesons can be classified as doublets of  $s_\ell^P$. Two states $(P,P^*)$ with $J^P=(0^-,1^-)$
correspond  to  $\ell=0$. The four states corresponding to  
$\ell=1$ can be collected in two  doublets, one  $(P^*_{0},P_{1}^\prime)$ with  $ s_\ell^P={1 \over 2}^+$  and  $J^P=(0^+,1^+)$,   another one $(P_{1},P_{2})$ with $ s_\ell^P={3 \over 2}^+$ and $J^P=(1^+,2^+)$. For $\ell=2$ the doublets have $s_\ell^P={3 \over 2}^-$ 
($(P^{*\prime}_1,P^*_2)$ with $J^P=(1^-,2^-)$) and $s_\ell^P={5\over2}^-$ ($(P^{*\prime}_2,P_3)$ with $J^P=(2^-,3^-)$).

In case of charm,   $m_c$ is greater than the strong interaction scale $\Lambda_{QCD}$ but it is not very large; therefore, 
corrections can be expected compared to the infinite limit. $O({1\over m_c})$ effects are the hyperfine splitting between mesons belonging to the same $s_\ell^P$ doublet,
and the  mixing of states with same $J^P$ and different  $s_\ell^P$,  namely the two axial vector states  with $s_\ell^P={1 \over 2}^+$ and ${3 \over 2}^+$.

The  six   $c \bar s$ states reported by  PDG 2006 \cite{PDG} can be classified according to this scheme.  $D_s$  and  $D^*_s$ belong to
 the  $ s_\ell^P={1 \over 2}^-$   doublet. 
There are   four  candidates  for the four $\ell=1$  states:   $D^*_{sJ}(2317)$ ($J^P=0^+)$, 
$D_{sJ}(2460)$ and  $D_{s1}(2536)$ ($J^P=1^+$), and  $D_{s2}(2573)$  ($J^P=2^+$).  The natural assignment is 
$D^*_{sJ}(2317)$  to the  $ s_\ell^P={1 \over 2}^+$  doublet and $D_{s2}(2573)$ to the  $ s_\ell^P={3 \over 2}^+$  doublet.
As for $D_{sJ}(2460)$ and  $D_{s1}(2536)$, they can be a mixing of the $1^+$ $ s_\ell^P={1 \over 2}^+$ and  ${3 \over 2}^+$  $c \bar s$
states. However, in  case of  non strange axial-vector $c \bar q$ mesons  the measured mixing angle  is small \cite{mixing}, a result confirmed by an analysis of $O({1\over m_c})$ effects breaking the heavy quark spin symmetry
\cite{Colangelo:2005gb}.   Invoking $SU(3)_F$, also the mixing angle in the case of $c \bar s$ is expected to be small, so that 
 $D_{s1}(2536)$ and $D_{sJ}(2460)$  essentially coincide with  the   $s_\ell^P={3 \over 2}^+$ and  ${1 \over 2}^+$ states.

In the above classification  $D_{sJ}(2860)$, which decays in two pseudoscalar mesons,  can be either  a $J^P=1^-$   $ s_\ell^P={3 \over 2}^-$ state,  or a
$J^P=3^-$   $ s_\ell^P={5 \over 2}^-$ state, i.e. 
a state with $\ell=2$ and  lowest radial quantum number.
Another possibility is that $D_{sJ}(2860)$ is a radial excitation of 
the  $J^P=1^-$   $ s_\ell^P={1 \over 2}^-$ state ($D_s^{*\prime}$),   of  the  $J^P=0^+$   $ s_\ell^P={1 \over 2}^+$ state  (first radial excitation of $D_{sJ}^*(2317)$) or of
the   $J^P=2^+$   $ s_\ell^P={3 \over 2}^+$ state ($D_{s2}^\prime$). 
The  $J^P$  assignment  can be done considering the decay modes and width. 

It was suggested\cite{Colangelo:2000jq}
 that a few high mass and high spin charm states  could be narrow
enough to be observed  and, in particular, that the $3^-$ state  belonging to the
$s_\ell^P={5\over 2}^-$  $c \bar q$ ($c \bar s$) doublet  is not too broad  since it decays to $D \pi$ ($D K$) in $f-$wave.   An analysis\cite{Colangelo:2006rq}  based on the heavy
quark limit \cite{positivep}  supports the assignment.  We define
 the fields representing the various heavy-light  meson doublets:    
$H_a$ for $s_\ell^P={1\over2}^-$ ($a$ light flavour index),
$S_a$ and $T_a$ for $s_\ell^P={1\over2}^+$ and $s_\ell^P={3\over2}^+$, respectively, and $X_a$ and $X^\prime_a$ for the doublets  corresponding  to  $\ell=2$, 
  $s_\ell^P={3\over2}^-$ and $s_\ell^P={5\over2}^-$, respectively:
\bea
H_a & =& \frac{1+{\rlap{v}/}}{2}[P_{a\mu}^*\gamma^\mu-P_a\gamma_5]  \label{neg}  \hspace*{1cm} ,  \hspace*{1cm} 
S_a = \frac{1+{\rlap{v}/}}{2} \left[P_{1a}^{\prime \mu}\gamma_\mu\gamma_5-P_{0a}^*\right]  \,\,\,, \nn \\
T_a^\mu &=&\frac{1+{\rlap{v}/}}{2} \left\{ P^{\mu\nu}_{2a} \gamma_\nu -P_{1a\nu} \sqrt{3 \over 2} \gamma_5 \left[
g^{\mu \nu}-{1 \over 3} \gamma^\nu (\gamma^\mu-v^\mu) \right]
\right\}  \,\,\, , \nn \\
X_a^\mu &=&\frac{1+{\rlap{v}/}}{2} \left\{ P^{*\mu\nu}_{2a} \gamma_5 \gamma_\nu -P^{*\prime}_{1a\nu} \sqrt{3 \over 2}  \left[
g^{\mu \nu}-{1 \over 3} \gamma^\nu (\gamma^\mu-v^\mu) \right]
\right\}   \,\,\,\, ,  \label{pos2}  \\
X_a^{\prime \mu\nu} &=&\frac{1+{\rlap{v}/}}{2} \left\{ P^{\mu\nu\sigma}_{3a} \gamma_\sigma -P^{*'\alpha\beta}_{2a} \sqrt{5 \over 3} \gamma_5 \left[
g^\mu_\alpha g^\nu_\beta -{1 \over 5} \gamma_\alpha g^\nu_\beta (\gamma^\mu-v^\mu)-  {1 \over 5} \gamma_\beta g^\mu_\alpha (\gamma^\nu-v^\nu) \right]
\right\}  \nn
\eea
with the various operators annihilating mesons of four-velocity $v$. The interaction of these particles with the
octet of light pseudoscalar mesons, introduced using 
 $\displaystyle \xi=e^{i {\cal M} \over
f_\pi}$, $\Sigma=\xi^2$,  the matrix ${\cal M}$ containing the octet of
$\pi, K$ and $\eta$ fields, and
$f_{\pi}=132 \; $ MeV,  is described by an effective Lagrangian invariant under chiral and
 heavy-quark spin-flavour transformations. At the leading order in the $1/m_Q$ and light meson momentum expansion,
the decays  $F \to H M$ $(F=H,S,T,X,X^\prime$ and $M$ a light pseudoscalar meson) are described by the Lagrangian  terms \cite{hqet_chir}: 
\bea
{\cal L}_H &=& \,  g \, Tr [{\bar H}_a H_b \gamma_\mu \gamma_5 {\cal
A}_{ba}^\mu ] \nn \\
{\cal L}_S &=& \,  h \, Tr [{\bar H}_a S_b \gamma_\mu \gamma_5 {\cal
A}_{ba}^\mu ]\, + \, h.c. \,, \nn \\
{\cal L}_T &=&  {h^\prime \over \Lambda_\chi}Tr[{\bar H}_a T^\mu_b
(i D_\mu {\spur {\cal A}}+i{\spur D} { \cal A}_\mu)_{ba} \gamma_5
] + h.c.   \label{lag-hprimo}   \\
{\cal L}_X &=&  {k^\prime \over \Lambda_\chi}Tr[{\bar H}_a X^\mu_b
(i D_\mu {\spur {\cal A}}+i{\spur D} { \cal A}_\mu)_{ba} \gamma_5
] + h.c.  \nn \\
{\cal L}_{X^\prime} &=&  {1 \over {\Lambda_\chi}^2}Tr[{\bar H}_a X^{\prime \mu \nu}_b
[k_1 \{D_\mu, D_\nu\} {\cal A}_\lambda+ k_2 (D_\mu D_\nu { \cal A}_\lambda + D_\nu D_\lambda { \cal A}_\mu)]_{ba}  \gamma^\lambda \gamma_5] + h.c.  \nn
\eea
where $\Lambda_\chi$ is  the chiral symmetry-breaking scale  ($\Lambda_\chi = 1 \, $ GeV), 
$D_{\mu ba}=-\delta_{ba}\partial_\mu+\frac{1}{2}\left(\xi^\dagger\partial_\mu \xi
+\xi\partial_\mu \xi^\dagger\right)_{ba}$ and
${\cal A}_{\mu ba}=\frac{i}{2}\left(\xi^\dagger\partial_\mu
\xi-\xi\partial_\mu \xi^\dagger\right)_{ba} $.
${\cal L}_S$ and ${\cal L}_T$ describe transitions of positive parity heavy mesons with
the emission of light pseudoscalar mesons in $s-$ and $d-$ wave, respectively,  with $g, h$ and $h^\prime$  effective coupling constants.  
${\cal L}_X$ and ${\cal L}_{X^\prime}$  describe the
transitions of higher mass mesons of negative parity with
the emission of light pseudoscalar mesons in $p-$ and $f-$ wave
with  couplings $k^\prime$, $k_1$ and $k_2$. 

At the same  order in the expansion in the light meson momentum,
the structure of the Lagrangian terms for radial excitations of  $H$, $S$ and $T$ doublets  does not
change,  but  the coupling constants $g, h$ and $h^\prime$ are  substituted by $\tilde g, \tilde h$ and $\tilde h^\prime$.  

In Table \ref{ratios} we collect  the ratios 
$\displaystyle {\Gamma( D_{sJ} (2860) \to D^*K) \over \Gamma( D_{sJ} (2860)\to DK) }$ 
and  $\displaystyle {\Gamma( D_{sJ} (2860)\to D_s \eta) \over \Gamma( D_{sJ} (2860)\to DK)  }$
obtained for various     quantum number assignments to $D_{sJ} (2860)$ \cite{Colangelo:2006rq}.
\begin{table}[ht]
    \caption{Predicted ratios
    $\displaystyle {\Gamma( D_{sJ} \to D^*K) \over \Gamma( D_{sJ} \to DK)
 }$   and   $\displaystyle {\Gamma( D_{sJ} \to D_s \eta) \over \Gamma( D_{sJ} \to DK)  }$ (with $DK=D^0K^+ + D^+ K_S^0$) for  various assignment
 of quantum numbers to  $D_{sJ}(2860)$.}
    \label{ratios}
    \begin{center}
    \begin{tabular}{| c | c | c | c |}
      \hline
 $D_{sJ}(2860) $  &  $D_{sJ}(2860) \to DK $&$\displaystyle{\Gamma( D_{sJ} \to D^*K) \over \Gamma( D_{sJ} \to DK)
 }$  &   $\displaystyle{\Gamma( D_{sJ} \to D_s \eta) \over \Gamma( D_{sJ} \to DK)  }$ 
\\ \hline
 $s_\ell^p={1\over 2}^-$, $J^P=1^-$,  rad. excit. & $p$-wave &$1.23$& $0.27$ \\
$s_\ell^p={1\over 2}^+$, $J^P=0^+$,  \hspace*{0.3cm} " & $s$-wave &$0$& $0.34$ \\
$s_\ell^p={3\over 2}^+$, $J^P=2^+$,  \hspace*{0.3cm}  " & $d$-wave &$0.63$& $0.19$\\
$s_\ell^p={3\over 2}^-$, $J^P=1^-$     \hspace*{0.6cm} & $p$-wave  & $0.06$& $0.23$ \\
$s_\ell^p={5\over 2}^-$, $J^P=3^-$     \hspace*{0.6cm}& $f$-wave  & $0.39$& $0.13$ \\
    \hline
    \end{tabular}
  \end{center}
\end{table}
These ratios   can be used to exclude some assignments. Indeed, since
 a $D^*K$ signal has not been observed (so far)  in the $D_{sJ}(2860)$ mass range,  
the production of $D^* K$ is not favoured and therefore  $D_{sJ}(2860)$ is  not a radial excitation of $D_s^*$ or  $D_{s2}$. 
The assignment 
$s_\ell^p={3\over 2}^-$, $J^P=1^-$ can also be excluded:   the  width  $\displaystyle
 \Gamma(D_{sJ}(2860)\to DK) $ obtained using   (\ref{lag-hprimo}) would be  $\Gamma(D_{sJ}(2860)\to DK)\ge 1$ GeV using $k^\prime \simeq h^\prime\simeq 0.45\pm0.05$  \cite{Colangelo:2005gb}, and there is no  reason  to presume that the coupling constant $k^\prime$ is sensibly smaller.
 
In the case of the assignment  $s_\ell^p={1\over 2}^+$, $J^P=0^+$, proposed   in some analyses,\cite{vanBeveren:2006st}
the decay  $D_{sJ}(2860)\to D^*K$ is forbidden and  the transition into $DK$  occurs 
in $s-$wave. The coupling costant for  the lowest radial quantum
number  was computed:  $h\simeq-0.55$
\cite{Colangelo:1995ph}; using this  value for $\tilde h$ we would obtain  $\Gamma(D_{sJ}(2860)\to DK)\ge1$ GeV.  It is reasonable to suppose that  $|\tilde h| < |h|$,  although no information is  available about  couplings of radially excited heavy-light mesons to low-lying states: the experimental width corresponds to  $\tilde h=0.1$.  A large signal   in the $D_s \eta$ channel would be expected.  
A problem is that,  if $D_{sJ}(2860)$ is a $0^+$ radial excitation, its  partner with 
 $J^P=1^+$  would decay to $D^* K$ with a width of the order of 40 MeV.
Since   both the lowest lying states  with $J^P= 0^+$ and $1^+$,
 $D^*_{sJ}(2317)$ and $D_{sJ}(2460)$,  are produced in charm continuum  at  $B$ factories,
to explain the absence of the $D^*K$ signal  at  energy around $2860$ MeV one must  invoke a mechanism favouring the production of the $0^+$ first radial excitation  and inhibiting the production of the $1^+$  radial excitation.

In the last case  $s_\ell^p={5\over 2}^-$, $J^P=3^-$
the narrow $DK$ width is due to the   kaon momentum  suppression:
$\displaystyle
\Gamma(D_{sJ}(2860)\to DK)\propto  q_K^7$.  A smaller but non negligible signal
in the $D^*K$ mode is predicted,  and  a small signal in the $D_s \eta$ mode is also expected. The  state of spin two  $D_{s2}^{*\prime}$ belonging to  the $s_\ell^P={5 \over 2}^-$ doublet, 
 which  can  decay to $D^* K$ and not to $DK$,
would  be narrow: $\Gamma(D_{s2}^{*\prime}\to D^*K)\simeq 50$ MeV for 
$m_Q\to \infty$: as an effect of $1/m_Q$ corrections,  $D_{s2}^{*\prime}\to D^*K$  can occur in $p$-wave,  in which case   $\Gamma(D_{s2}^{*\prime})$ could be  broader.

 $D_{sJ}(2860)$ with  $J^P=3^-$ is not expected to  be produced
 in non leptonic $B$ decays such as 
$B^0 \to  D^- D_{sJ}(2860)^+$ and $B^+ \to  \bar D^0 D_{sJ}(2860)^+$: 
indeed in the   Dalitz plot analysis of $B^+ \to \bar D^0 D^0 K^+$  Belle found no signal of $D_{sJ}(2860)$    \cite{belle2715}.
The non-strange partner $D_3$ of a $J^P=3^-$ $D_{sJ}(2860)$ state,     if the mass splitting
$M_{D_{sJ}(2860)}-M_{D_3}$ is of the order of the strange quark mass, is  also expected to be narrow:
$\Gamma(D_{3}^+\to D^0 \pi^+ )\simeq 37$ MeV. It  can  be 
produced in semileptonic as well as in non leptonic $B$ decays,
such as $B^0 \to D_3^- \ell^+  \bar \nu_\ell$ and $B^0 \to D_3^- \pi^+$. \cite{Colangelo:2006rq} 

\begin{figure}[tbh]
\centering
\includegraphics[scale=0.45] {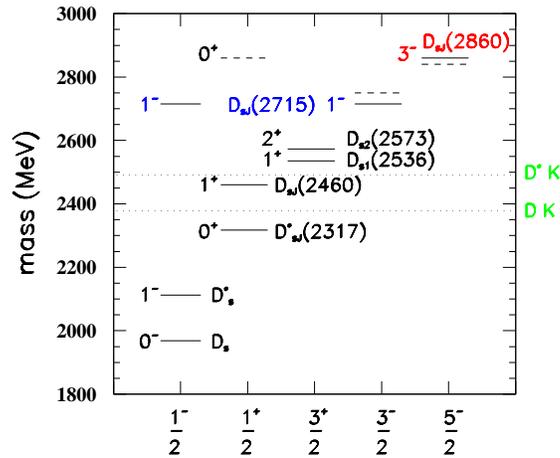} 
\vspace*{-1cm}
\caption{\footnotesize{ $c \bar s$ spectrum with possible assignments of $D_{sJ}(2860)$ 
and $D_{sJ}(2715)$. 
}}\label{fig:spectrumcs}
\end{figure}

The analysis of $D_{sJ}(2715)$  can be done analogously and  is in progress. A  proposal for the $c \bar s$ spectrum is shown in fig.\ref{fig:spectrumcs}.
%

\section{Hidden charm mesons and  $X(3872)$}

A puzzle in the hidden charm sector is  the  meson $X(3872)$ discovered in the $J/\psi \pi^+ \pi^-$  invariant mass distribution  in  $B$
decays   and in $p \bar p$ collisions  \cite{Choi:2003ue}, with  $M(X)=3871.2 \pm
0.5$ MeV  and  $\Gamma(X)<2.3$ MeV ($90\%$ C.L.).\cite{PDG}
The
$\pi^+ \pi^-$ spectrum  is peaked  for large invariant mass.\cite{pipispectrum}
$X(3872)$  was
not   observed in $e^+ e^-$ annihilation and in $\gamma \gamma$ fusion; 
searches of  charged partners  also produced negative results. 
The  charge conjugation of the state is C=+1  since the   mode $X \to J/\psi \gamma$ was observed
\cite{belle3p}; angular distribution studies show that  the most  likely  quantum number assignment 
is $J^{PC}=1^{++}$.\cite{Abe:2005iy}

 Furthermore, a  near-threshold  $D^0 \bar D^0 \pi^0$  enhancement in
 $B \to D^0 \bar D^0 \pi^0 K$ decay
was recently reported, with the peak at $M=3875.4 \pm 0.7 ^{+1.2}_{-2.0}$ MeV and
$B(B\to K X \to K D^0 \bar D^0  \pi^0)=(1.27 \pm 0.31^{+0.22}_{-0.39}) \times 10^{-4}$
\cite{Gokhroo:2006bt}.
If the enhancement is only due to $X(3872)$ one finds 
  $\frac{B(X \to D^0 \bar D^0  \pi^0)}{B(X \to J/\psi \pi^+ \pi^- )}=9\pm4$,
hence $X$  mainly  decays into final states with open charm mesons.
Notice that the central value of the mass measured in 
  $D^0 \bar D^0 \pi^0$   is $4$ MeV   higher than the PDG value (with a large  systematic error).

Since another hadronic  decay mode was observed for  $X(3872)$:
 $X \to J/\psi \pi^+ \pi^- \pi^0$  with  $\frac{B(X \to J/\psi \pi^+ \pi^- \pi^0)}{B(X \to J/\psi \pi^+ \pi^- )}=1.0 \pm 0.4 \pm 0.3$ \cite{belle3p,Y3930}, there are G-parity violating $X$ transitions  or, if the
 two modes are
 considered as  induced by $\rho^0$ and $\omega$ intermediate states, isospin violation:
 this  suggested the conjecture  that  $X(3872)$ is not a charmonium $\bar c c$ state. In the search of  the right interpretation,  the   coincidence between the $X$ mass  as 
 averaged by PDG   and the $D^{*0} \overline D^0$ mass: 
 $M(D^{*0} \overline D^0)=3871.2\pm 1.0$ MeV,   inspired  the proposal that
$X(3872)$  could be  a realization of the molecular quarkonium\cite{okun}, a  $D^{*0}$ and $\overline D^0$ bound state with small binding energy \cite{molec},  an interpretation that would allow to account for a few properties of $X(3872)$. For example,
 describing the wave function of $X(3872)$  through various hadronic components \cite{voloshin1}:
\be
|X(3872)>=a \, |D^{*0} \bar D^0+ \bar D^{*0}  D^0> + b \,  |D^{*+}  D^-+  D^{*-}  D^+> + \dots
\ee
(with $|b| \ll |a|$)
one could explain why this state seems not to have  definite isospin,  why
 the mode $X \to J/\psi \pi^0 \pi^0$ was not  found, and why, if the molecular binding
 mechanism is provided by a single pion exchange, there are no  $D \overline D$ molecular states.
 It has also been suggested that the molecular interpretation   implies that
 the radiative decay in neutral $D$ mesons: $X \to D^0 \bar D^0 \gamma$ should be dominant with respect to $X \to D^+  D^- \gamma$ \cite{voloshin1}.

The description of $X(3872)$ in a simple charmonium  scheme,  in which  it  would be identified  as the first radial excitation of the $J^{PC}=1^{++}$ state,
  presents alternative arguments  to the molecular description\cite{charmonium}.  For example,
  the molecular binding mechanism  has  not been clearly identified.\cite{suzuki,reviews}
Concerning  the large value of the ratio
$\frac{B(X \to J/\psi \pi^+ \pi^- \pi^0)}{B(X \to J/\psi \pi^+ \pi^- )}$ one has to
consider that phase space effects in two and three pion modes are very different.
The ratio of the amplitudes is  smaller:
$ \frac{A(X \to J/\psi \rho^0)}{A(X \to J/\psi \omega)}\simeq 0.2$, so that the
isospin violating amplitude is
20\% of the isospin conserving one, an
 effect that could be related to another isospin violating effect,  the mass difference between
neutral and charged $D$ mesons, considering  the contribution of  $DD^*$ intermediate states
to $X$ decays.  The  prediction
$\Gamma(B^0 \to X K^0) \simeq \Gamma(B^- \to X K^-)$,   based on  the charmonium description, is neither confirmed nor excluded, since
 $ \frac{B(B^0 \to K^0 X)}{B(B^+ \to K^+ X)}=0.50 \pm 0.30 \pm 0.05$\cite{babarkpi}.
The $\bar c c$ interpretation leaves unsolved  the issue of the
eventual overpopulation of the level corresponding to  the first radial excitations of $1^{++}$ $\bar c c$ states resulting from the possible assignment of these quantum numbers to another structure  observed by Belle Collaboration,
$Y(3930)$ \cite{Y3930},  however, since this other resonance is still not  confirmed and its properties not
fully understood, 
 the charmonium option for $X(3872)$ seems  not excluded, yet.  A warning comes from the
 $D^0 \bar D^0 \pi^0$ signal which can contribute to settle the question of
 the coincidence of the $X$ and $D^0 \bar D^{*0}$ mass: an $X$ mass above the $D^0 \bar D^{*0}$ 
 threshold would be difficult to explain in the molecular scheme.\cite{cusp}
 
\begin{figure}[t]
\begin{center}
\includegraphics[scale=0.6]{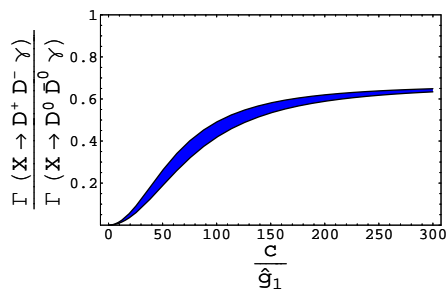}
\end{center}
\caption{\footnotesize{Ratio of charged $X \to D^+ D^- \gamma$ to
neutral $X \to D^0 \bar D^0 \gamma$ decay  widths  versus $c/\hat g_1$.}}\label{fig:ratio} 
\end{figure}

  The suggestion   that  observation of the dominance
of  $X \to D^0 \bar D^0 \gamma$  with respect to $X \to
D^+  D^- \gamma$ can be interpreted as a signature
of the molecular structure of $X(3872)$\cite{voloshin1}  is also problematic.\cite{Colangelo:2007ph}
Assuming that $X(3872)$ is an ordinary $J^{PC}=1^{++}$ charmonium
state together with  a standard
mechanism for $X$ radiative transition into charmed mesons,  the
ratio $R={\Gamma(X \to D^+ D^- \gamma)\over \Gamma(X
\to D^0 \overline D^0 \gamma)}$ is small, and   it  is tiny in a wide
 range of the hadronic parameters governing the
decays,  therefore  $R\ll 1$ is  not peculiar of 
  a molecular quarkonium $X(3872)$.\cite{Colangelo:2007ph}
%
\begin{figure}[b]
\begin{center} 
\includegraphics[width=0.32\textwidth] {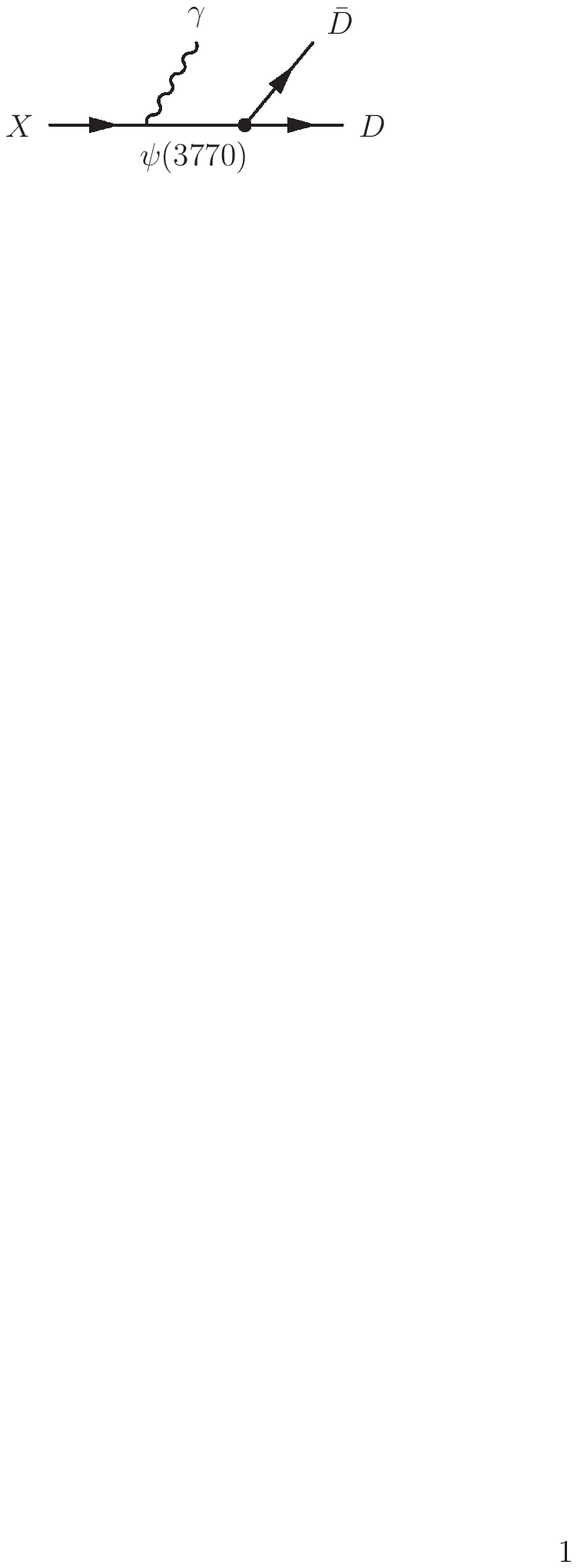}
\includegraphics[width=0.32\textwidth] {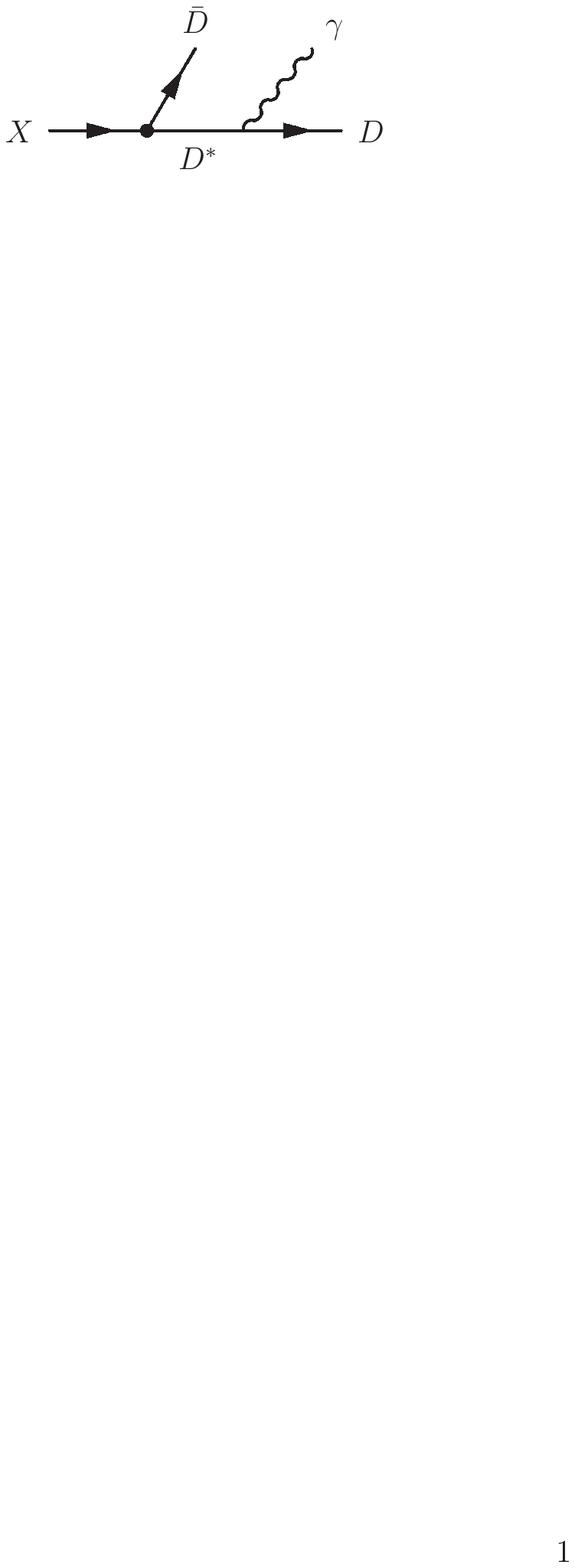}
\includegraphics[width=0.32\textwidth] {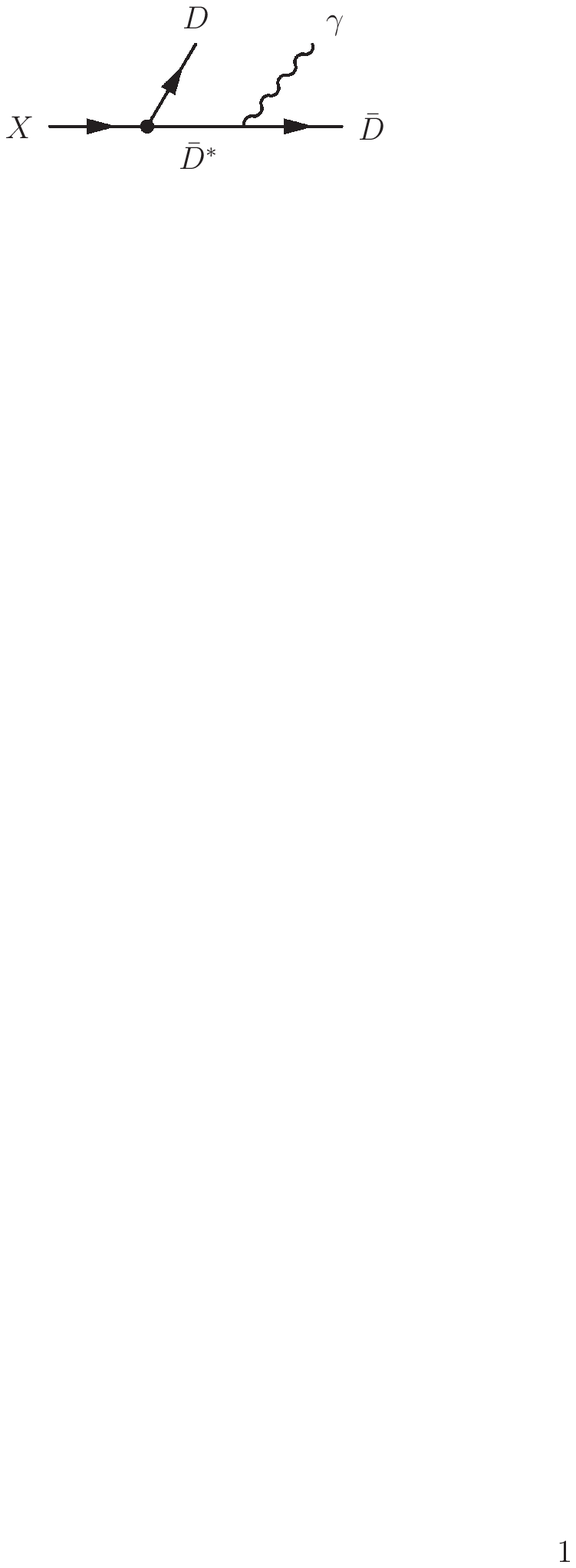}
 \vspace*{-11cm}
 \caption{Diagrams contributing to  $X(3872) \to D \bar D \gamma$.}
  \label{fig:gamma}
 \end{center}
\end{figure}
%
%
This can be demonstrated describing  the   $X(3872)\to D \bar D \gamma$ amplitude by  diagrams with
intermediate particles nearest to their mass shell, as those
depicted in fig.\ref{fig:gamma} with  $D^*$  and  $\psi(3770)$   as intermediate states. The
amplitude can be expressed in terms of two  unknown  quantities:
a coupling $\hat g_1$ governing the $X \bar D D^*(D \bar D^*)$ matrix elements,  and a
 coupling $c$ appearing in the $X \psi(3770) \gamma$ matrix element, 
all the other quantities being fixed by
 experimental data.\cite{pham,Colangelo:2007ph}. As shown  in fig.\ref{fig:ratio},  the ratio $R$
is tiny for small values of $\displaystyle {c / \hat g_1}$,  

\begin{figure}[h]
\begin{center}
\includegraphics[width=0.3\textwidth] {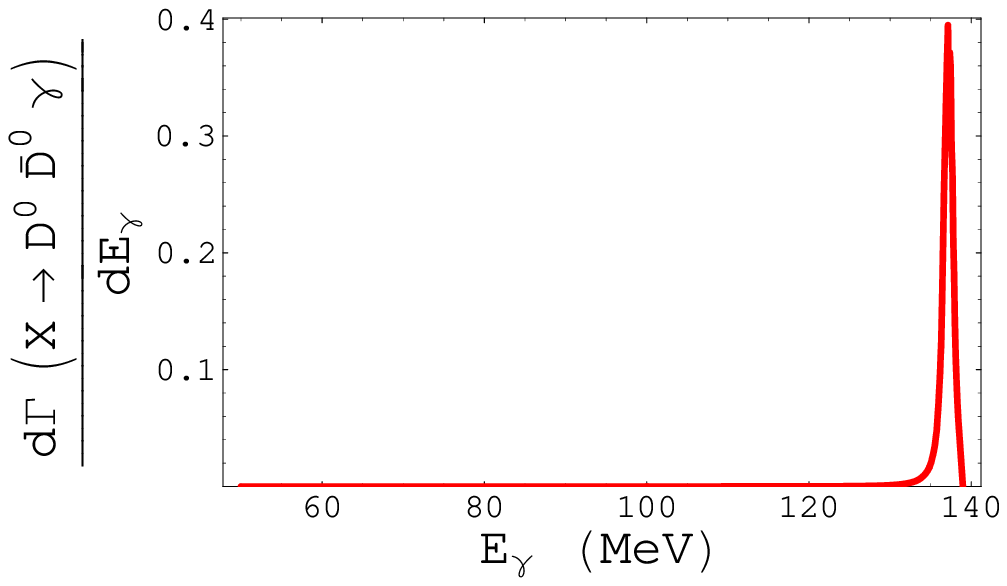} \hspace{0.3cm}
 \includegraphics[width=0.3\textwidth] {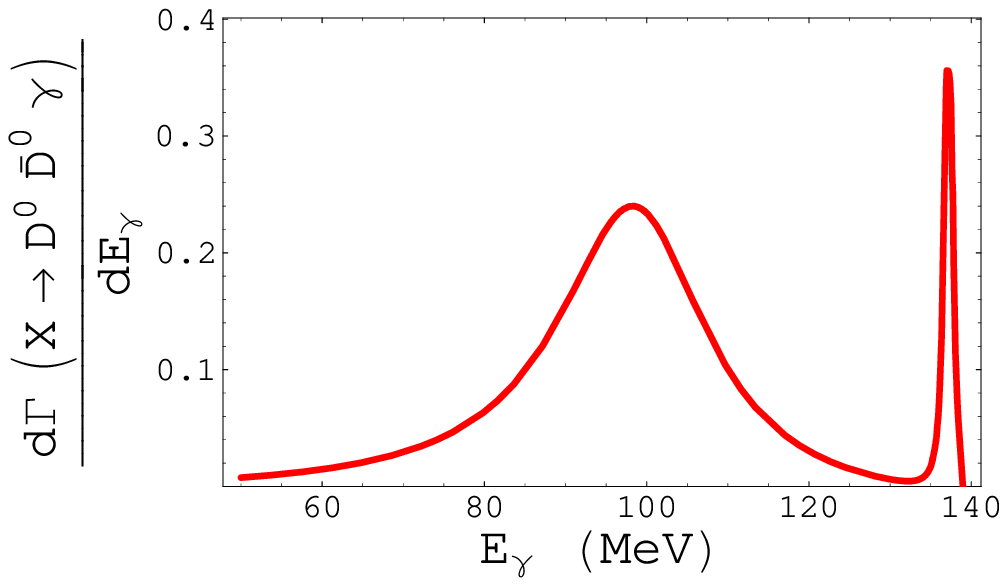}\\ \vspace{0.3cm}
 \includegraphics[width=0.3\textwidth] {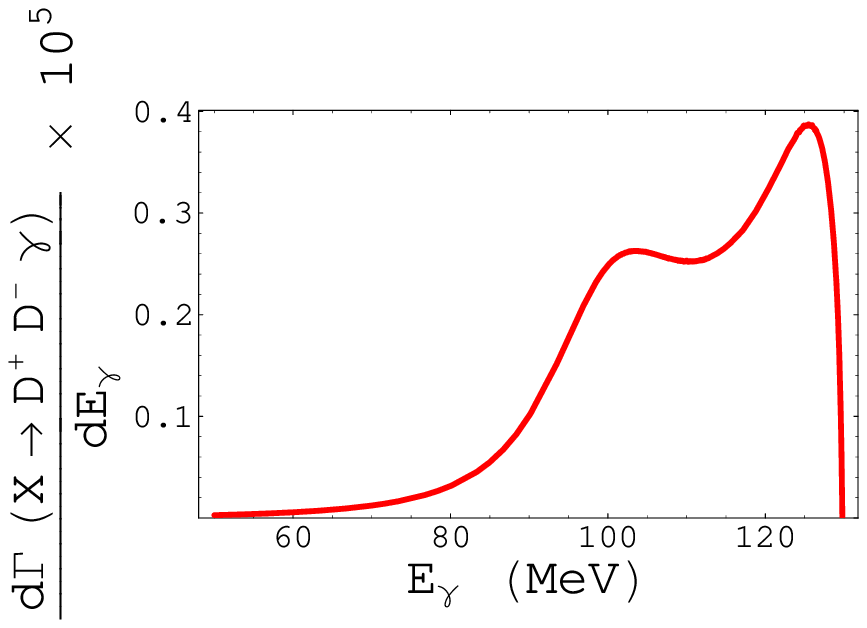} \hspace{0.3cm}
 \includegraphics[width=0.3\textwidth] {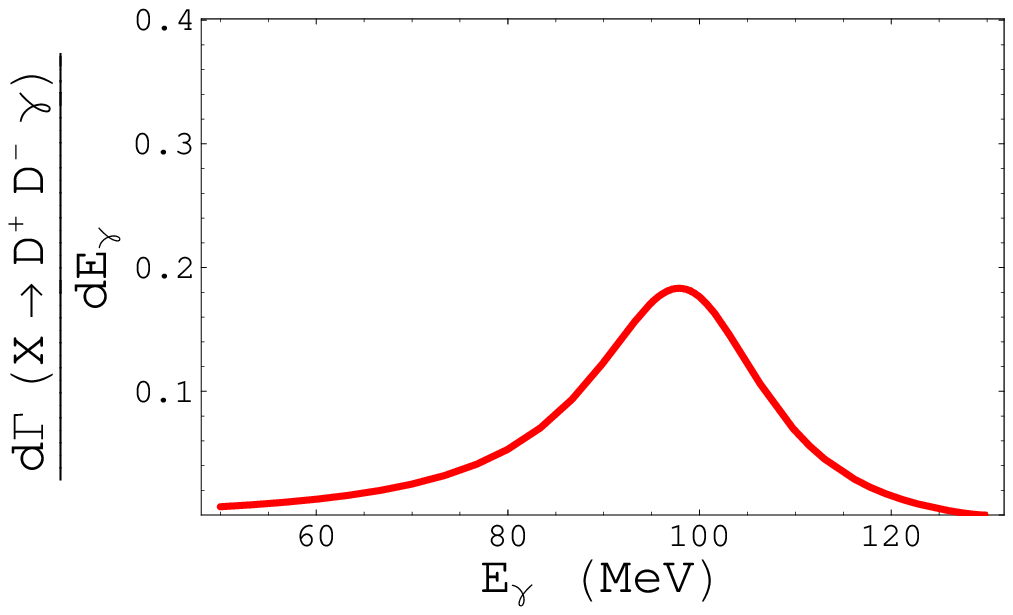}
\end{center}
\caption{\footnotesize{Photon spectrum (in arbitrary units) in
$X \to D^0 \bar D^0 \gamma$ (top) and  $X \to D^+ D^- \gamma$ (bottom)
decays  for values of the hadronic parameter $c/\hat g_1=1$ (left) and $c/\hat g_1=300$ (right).}}\label{fig:spectra} 
\end{figure}

The photon spectrum is different in case of a charmonium or a molecule. It is interesting to consider it in  $X$ decays to neutral and charged
$D$ meson pairs for two representative values:  $\displaystyle{c /\hat g_1}=1$ and $300$
 (fig.\ref{fig:spectra}).
For low value of $\displaystyle{c/ \hat  g_1}$, i.e. in the condition where the intermediate $D^*$ dominates the decay amplitude, the photon spectrum in the $D^0 \bar D^0 \gamma$ mode  
coincides with the line corresponding to the $D^*$ decay at $E_\gamma \simeq 139$ MeV. The narrow peak is different from the line shape expected
in a molecular description, which is related to the wave function of the two heavy mesons bounded in the $X(3872)$,  in particular  to the binding energy of the system, being broader for larger binding energy. On the other hand, the photon spectrum in the charged $D^+  D^- \gamma$ mode is broader,
with a peak at  $E_\gamma \simeq 125$ MeV, the total  $X \to D^+  D^- \gamma$ rate being severely suppressed with respect to    $X \to D^0  \bar D^0 \gamma$.

At the opposite side of the $\displaystyle{ c / \hat g_1}$  range,  where $\psi(3770)$ gives a large contribution to the radiative amplitude, a peak at $E_\gamma \simeq 100$ MeV appears  both in neutral and charged
$D$ meson modes, in the first case together with  the structure at  $E_\gamma \simeq 139$ MeV. This spectrum was previously  described   and the radiative decay was interpreted as due to the $\bar c c$ core of  $X(3872)$\cite{voloshin1}. 
So, the measurement  of the photon spectrum $\Gamma(X \to D \bar D \gamma)$ could be used
 to  shed light on the structure of $X(3872)$.

\section{Conclusions}
A few  results in charm spectroscopy challenge our understanding. More than thirty years after the first observation, charm continues to be a surprise for us.

\section*{Acknowledgements}
One of the authors (PC)  thanks the Yukawa Institute for Theoretical Physics at Kyoto University 
and the organizers of the  Workshop YKIS2006 on  "New Frontiers on QCD"  for kind hospitality. Work supported in part by the EU Contract  No. MRTN-CT-2006-035482, "FLAVIAnet".

%


\end{document}